\documentclass{caosp}
\usepackage{graphicx}

\articleNo{C04}
\pubyear{2014}
\volume{43}
\volnumber{3}
\firstpage{1}
\received{October 31, 2013}
\accepted{January 30, 2014}

\begin{document}

\hauthor{R.\,Errmann, St.\,Raetz, M.\,Kitze, R.\,Neuh{\"a}user and the YETI
team}

\htitle{The search for transiting planets using the YETI network}
\title{The search for transiting planets using the YETI 
       network\thanks{* Some of the data presented herein were obtained 
                      at the W.M. Keck Observatory.}}
\renewcommand\footnotemark{$^*$}

\author{
        R.\,Errmann\inst{1,2}
      \and St.\,Raetz\inst{1}
      \and M.\,Kitze\inst{1}
      \and R.\,Neuh{\"a}user\inst{1}
      \and the YETI team
       }

\institute{
           Astrophysikalisches Institut und Universit\"ats-Sternwarte, Schillerg\"a{\ss}chen 2-3, 07745 Jena, Germany, \email{ronny.errmann@uni-jena.de}
         \and 
           Abbe Center of Photonics, Friedrich-Schiller-Universit\"at Jena, Max-Wien-Platz 1, 07743 Jena, Germany
          }

\date{October 31, 2003}

\maketitle

\begin{abstract}
To search for young transiting planets in continuous light curves, we
monitor young open clusters (2-200\,Myr) with the YETI network.  Here we
report the first transiting candidates (two in Trumpler\,37, one in
25\,Ori).  Follow-up observations of the candidates are partly done.
\end{abstract}

%
\section{Introduction}

\label{intr}
\begin{figure}[t]
\centerline{\includegraphics[width=0.98\textwidth,clip=]{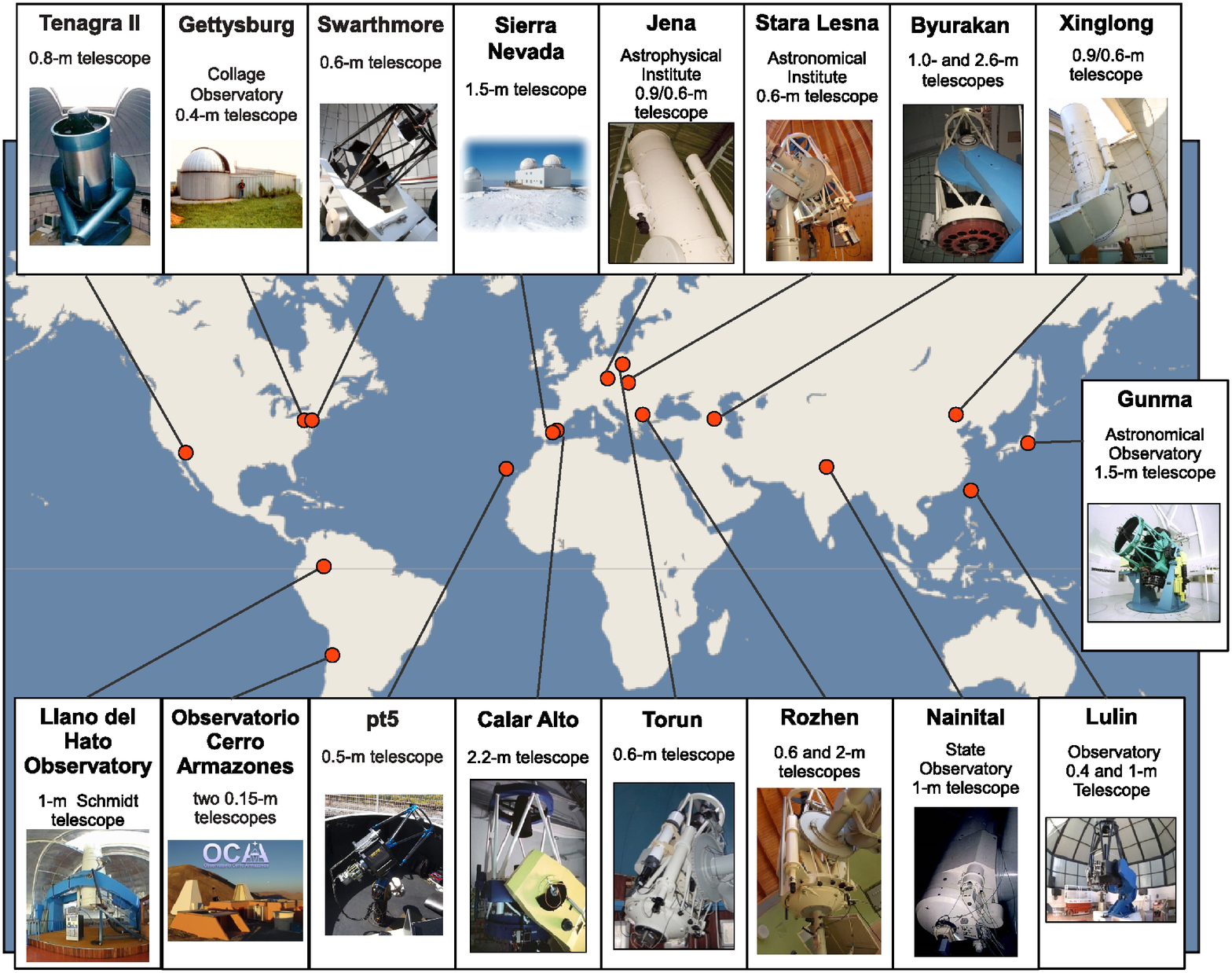}}
\caption{Map of the YETI telescopes. With the network it is possible to observe continuously.}
\label{Fig:YETI_map}
\vspace*{-5.mm}
\end{figure}

Young transiting planets could play a key role to distinguish planet formation scenarios, as it is possible to test evolutionary models with the well determined parameters of such planets, like radius, mass, and age.
As no transiting planets with ages younger than $100$\,Myr were found so far, we monitor young open clusters to search for transit signals among young stars.
To increase phase coverage, continuous observations are needed. Therefore we established YETI (Young Exoplanet Transit Initiative), a network of ground-based telescopes with mirror diameters of $0.4$ to $2$\,m (see Neuh{\"a}user {\it et al.} \cite{neu11} for further details). Fig.\,\ref{Fig:YETI_map} shows a map with all telescope sites as well as the telescope sizes. Each cluster is observed for at least three years with three campaign runs of lengths longer than one week. Observations are done in the $R$-band.

The first two target clusters of YETI were Trumpler\,37 (Tr37) and 25\,Ori. Some properties of both clusters are summarized in Tab.\,\ref{Tab:cluster_prop}.
The monitoring of Tr37 started in summer 2009 at the University Observatory Jena. Data of the YETI telescopes were gathered in summers 2010 and 2011, each year three campaign runs were performed. 
The observations of 25\,Ori at the University Observatory Jena started in January 2010. In this first phase of the observation  two additional telescopes (Gunma/Japan and CIDA/Venezuela) joined the photometric monitoring in January and February 2010.
25\,Ori became a target of the YETI project in the winter seasons 2010/2011, 2011/2012, and 2012/2013. Fifteen different observatories, spread worldwide at different longitudes, participated in the campaigns. 

Data reduction (dark/bias and flat-field correction) as well as aperture photometry with adjusted aperture to the seeing conditions was done night by night and telescope by telescope. For a particular star, the final lightcurve was created by doing differential photometry (Broeg {\it et al.} \cite{bro05}) for all nights and telescopes on a small sample of comparison stars (similar brightness and color, small angular distance). A detailed description can be found in Errmann {\it et al.} (\cite{err13b}).

\begin{table}
\small
\begin{center}
\caption{Properties of the stars in the clusters Trumpler\,37 and 25\,Ori.}
\label{Tab:cluster_prop}
\vspace*{-3.mm}
\begin{tabular}{lcc}
\hline\hline
                    & Trumpler\,37      &  25\,Ori      \\
\hline
Age [Myr]           & $\sim4$ [1]       & 7-10 [2]      \\
Distance [pc]       & $\sim870$ [3]     & $\sim330$ [4] \\
Number member stars & 774 [5]           & $\sim250$ [6] \\
\hline\hline
\multicolumn{3}{c}{
[1]~Kun {\it et al.} (\cite{kkb08}); 
[2]~Brice{\~n}o {\it et al.} (\cite{bri07});
[3]~Contreras {\it et al.} (\cite{con02});
}\\
\multicolumn{3}{c}{
[4]~Brice{\~n}o {\it et al.} (\cite{bri05});
[5]~Errmann {\it et al.} (\cite{err13a});  
[6]~Brice{\~n}o {\it et al.} (\cite{bri13})
}
\end{tabular}
\end{center}
\vspace*{-5.mm}
\end{table}

\begin{figure}
\centerline{\includegraphics[height=0.93\textwidth,angle=270]{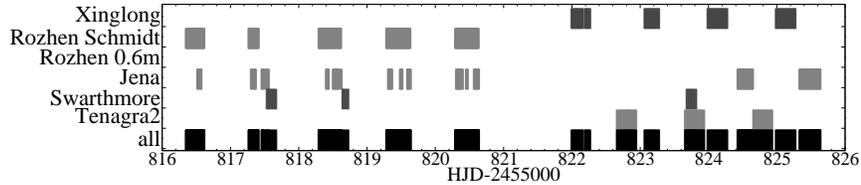}}
\caption{Observation times of the first transiting candidate in Trumpler\,37 during the third YETI campaign in 2011.}
\label{Fig:timecoverage}
\vspace*{-5.mm}
\end{figure}

\begin{figure}
\centerline{\includegraphics[height=0.93\textwidth,angle=270]{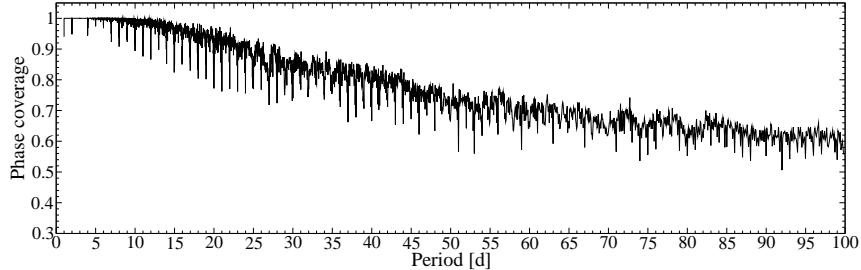}}
\caption{Phase coverage of the first transiting candidate in Trumpler\,37.}
\label{Fig:phasecoverage}
\vspace*{-5.mm}
\end{figure}

Fig.\,\ref{Fig:timecoverage} shows the time coverage of the observations for a star in Tr37 for the third YETI campaign in 2011. As the star is located in a larger separation to the cluster center, it does not fit into the field of view of some telescopes. Furthermore, not all telescopes could always allocate time and some suffer from seasonal bad weather, hence only 5 of the YETI sites are present. But even with that smaller number of telescopes, we could observe nearly continuously for $48.5$\,h at the end of the campaign. Fig.\,\ref{Fig:phasecoverage} shows the phase coverage of the same star. 
 For a period up to 10\,d we reach 100\% phase coverage and even for a period of 50\,d the phase coverage is better than 70\%. The coverage at periods of a multiple of a full day is slightly worse, as a telescope in the pacific ocean is missing.

\section{Transiting candidates}
In the Tr37 data two transit candidates were found using a simple box fitting algorithm. Another transiting candidate, first reported in van Eyken {\it et al.} (\cite{eyk12}), is also visible in the 25\,Ori data from YETI. 

\subsection{First candidate in Trumpler\,37}
At the late F-type star ($R=15.1$\,mag), a transit-like signal of $\Delta R=54$\,mmag and Period $P=1.36491$\,d is visible. The time and phase coverage for this star is shown in Fig.\,\ref{Fig:timecoverage} and Fig.\,\ref{Fig:phasecoverage}, respectively.

\begin{figure}
 \centerline{\includegraphics[height=\textwidth,angle=270]{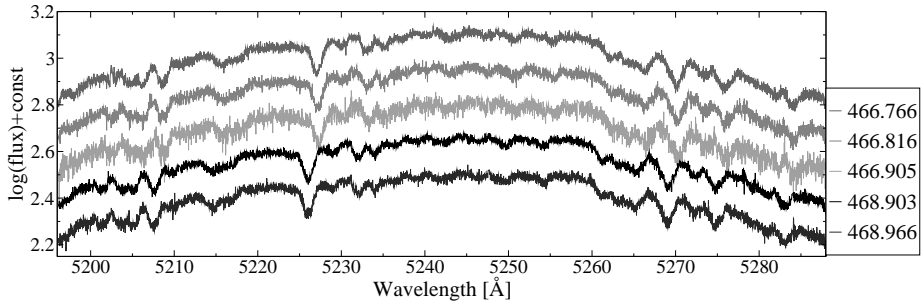}}
  \caption{All obtained Keck spectra of the fourth order of the middle chip, sorted by time by adding adapted offsets (time is given in JD-2455000). The flux per spectrum is $\sim300$\,ADU. Between the spectra of the two nights the shifts of the lines are visible.}
 \label{Fig:Keck-spec}
\vspace*{-5.mm}
\end{figure}

The follow-up observations of this candidate are already finished and will be presented in detail in Errmann {\it et al.} (\cite{err13b}). We got a high-quality light curve in the $I$-band and a low-resolution spectrum from \textit{CAFOS} at the 2.2\,m telescope at Calar Alto. A medium-resolution spectrum was taken with \textit{Hectochelle} at MMT. The host star has a late F spectral type. Eclipsing background stars in the optical point-spread function could be ruled out from high-resolution images from IRCS observations at Subaru. To solve the radial velocity orbit, 5 high-resolution spectra were observed with \textit{HIRES} at Keck-I. As the latter observations were done near the quadratures, the line shifts are already visible by eye in Fig.~\ref{Fig:Keck-spec}. The radial velocity variations of $\sim35$\,km/s semi-amplitude are too large for a sub-stellar mass companion, hence we found a false positive. The mass of the companion is about 0.15 to 0.35\,M$_{\sun}$.

\subsection{Second candidate in Trumpler\,37}
The $R=13.4$\,mag bright star shows a transit-like lightcurve with a period of $P\sim0.74$\,d and depth of $\Delta R=11$\,mmag. Additional, semi-periodic brightness variations of $\Delta R\sim15$\,mmag over time scale of $9-10$\,d are visible. We are actively working on the follow-up observations of this candidate.

\subsection{Transit candidate in 25\,Ori}
\begin{figure}
\centerline{\includegraphics[height=0.95\textwidth,angle=270]{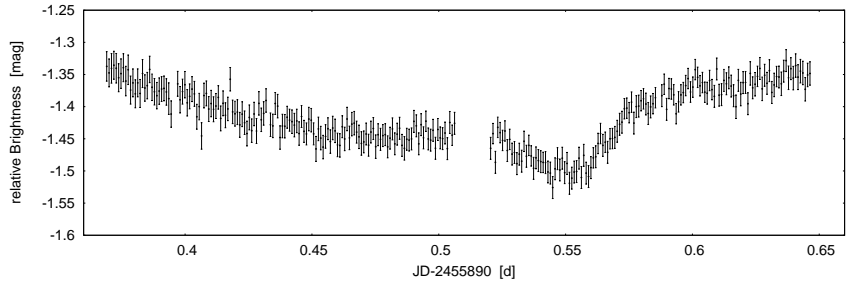}}
\caption{Lightcurve of the transiting candidate in 25 Ori in a single night. The transit occurs between 0.52 and 0.58. As the host star is a weak-line T~Tauri star, variations outside the transit are visible.}
\label{Fig:lc_25Ori}
\vspace*{-5.mm}
\end{figure}

The host star (CVSO30) is a weak-lined T~Tauri star and therefore very active. This causes brightness variations up to $200$\,mmag, while the depth of the transit is only $\sim40$\,mmag (see Fig.\,\ref{Fig:lc_25Ori}).
An interesting feature of the transit light curve of CVSO30 was mentioned by van Eyken {\it et al.} (\cite{eyk12}). They observed two sets of light curves in the years 2009 and 2010. It can clearly be seen that there is an overall change in the transit shape between the two years data sets, which could be explained by gravity darkening of the host star (Barnes {\it et al.}, \cite{bar13}). Further photometric follow-up can test the proposed spin-orbit misalignment.

\section{Outlook}
The monitoring of the clusters Trumpler\,37 and 25\,Ori with YETI has already finished, while the monitoring of IC\,348, Collinder\,69, NGC\,1980, and NGC\,7243 is ongoing. It is expected to find more transiting candidates, as we improve our routines and do further data analysis.

\acknowledgements
We thank all of the participating YETI Observatories for the observations.
All the participating observatories appreciate the logistic and financial
support of their institutions and in particular their technical workshops.
RN, MK, RE, and SR would like to thank DFG for support in the Priority Programme SPP 1385 on the {\em First ten Million years of the Solar System} in projects NE 515 / 33-1/2 and 34-1/2.
We would like to acknowledge financial support from the Thuringian government (B 515-07010) for the STK CCD camera used in this project.

\end{document}